%% file: ai4s-IEEE.tex
\documentclass[conference]{IEEEtran}
\IEEEoverridecommandlockouts
\usepackage{cite}
\usepackage{amsmath,amssymb,amsfonts}
\usepackage{algorithmic}
\usepackage{graphicx}
\usepackage{textcomp}
\usepackage{xcolor}
\usepackage{url}

\RequirePackage{color}
\usepackage{colortbl}
\definecolor{RED}{rgb}{1,0,0}
\definecolor{BLUE}{rgb}{0,0,1}
\definecolor{GREEN}{rgb}{0,1,0}
\definecolor{DGREEN}{RGB}{0,140,51}
\definecolor{color1}{rgb}{0.913, 0.776, 0.686}
\definecolor{color2}{rgb}{0.913, 0.867, 0.686}
\definecolor{ltgray}{rgb}{0.85, 0.85, 0.85}

\usepackage{tcolorbox}
\usepackage{colortbl}
\usepackage{mdframed}

\newenvironment{myenv}[1]
  {\mdfsetup{
    frametitle={\colorbox{white}{\space#1\space}},
    innertopmargin=1pt,
    frametitleaboveskip=-\ht\strutbox,
    frametitlealignment=\center
    }
  \begin{mdframed}%
  }
  {\end{mdframed}}

\usepackage{listings}
\usepackage{stackengine}

\usepackage[tight,footnotesize]{subfigure}

\newcommand{\bi}{\begin{itemize}}
\newcommand{\ei}{\end{itemize}}
\newcommand{\be}{\begin{enumerate}}
\newcommand{\ee}{\end{enumerate}}
\newcommand{\im}{\item}

\def\BibTeX{{\rm B\kern-.05em{\sc i\kern-.025em b}\kern-.08em
    T\kern-.1667em\lower.7ex\hbox{E}\kern-.125emX}}
\begin{document}

\title{ChatVis: Automating Scientific Visualization with a Large Language Model}

\author{\IEEEauthorblockN{
%1\textsuperscript{st} 
Tanwi Mallick}
\IEEEauthorblockA{
%\textit{dept. name of organization (of Aff.)} \\
\textit{Argonne National Laboratory}\\
Lemont, IL, USA \\
tmallick@anl.gov}
\and
\IEEEauthorblockN{
%2\textsuperscript{nd} 
Orcun Yildiz}
\IEEEauthorblockA{
%\textit{dept. name of organization (of Aff.)} \\
\textit{Argonne National Laboratory}\\
Lemont, IL, USA \\
oyildiz@anl.gov}
\and
\IEEEauthorblockN{
%2\textsuperscript{nd} 
David Lenz}
\IEEEauthorblockA{
%\textit{dept. name of organization (of Aff.)} \\
\textit{Argonne National Laboratory}\\
Lemont, IL, USA \\
dlenz@anl.gov}
\and
\IEEEauthorblockN{
%3\textsuperscript{rd} 
Tom Peterka}
\IEEEauthorblockA{
%\textit{dept. name of organization (of Aff.)} \\
\textit{Argonne National Laboratory}\\
Lemont, IL, USA \\
tpeterka@mcs.anl.gov}
}

\maketitle

\begin{abstract}
We develop an iterative assistant we call ``ChatVis'' that can synthetically generate Python scripts for data analysis and visualization using a large language model (LLM). The assistant allows a user to specify the operations in natural language, attempting to generate a Python script for the desired operations, prompting the LLM to revise the script as needed until it executes correctly. The iterations include an error detection and correction mechanism that extracts error messages from the execution of the script and subsequently prompts LLM to correct the error. Our method demonstrates correct execution on five canonical visualization scenarios, comparing results with ground truth. 
We also compared our results with scripts generated by several other LLMs without any assistance. In every instance, ChatVis successfully generated the correct script, whereas the unassisted LLMs failed to do so. The code is available on GitHub: \url{https://github.com/tanwimallick/ChatVis/}.
\end{abstract}


\section{Introduction}\label{sec:Intro}

\input{Intro}

\section{Related Work}\label{sec:Related}

\input{Related}

\section{Methodology}
\label{sec:Methodology}
\input{Methodology}

\section{Experimental Results}
\label{sec:Results}
\input{Results}

\section{Conclusion}
\label{sec:Conclusion}
\input{Conclusion}

\section{Acknowledgments} 
This material is based upon work supported by the U.S. Department of Energy, Office of Science, Office of Advanced Scientific Computing Research, under contract numbers DE-AC02-06CH11357, program manager Margaret Lentz.

\bibliographystyle{plain}
\bibliography{references}

\end{document}

%% file: Intro.tex
Scientific visualization is typically performed one of two ways: interacting through a graphical user interface with a visualization tool such as 
ParaView~\cite{ayachit15} or VisIt~\cite{childs12},
or writing an offline script---usually in Python---for the same tool. Either way,
the visualization is created manually, through trial and error, one step at a time. Expert knowledge of data analysis and visualization is required, and the resulting visualizations are time-consuming to create and difficult to reproduce.

We propose a new approach to address the problems of productivity and usability of creating data analysis and visualization through synthetic software generation using an LLM. We develop an iterative prompt assistant, ChatVis, that allows the user to specify a chain of analysis/visualization operations in natural language. The assistant attempts to generate a Python script for the desired operations and iterates until the script executes correctly, prompting the LLM to revise the script as needed. We use ParaView as the visualization tool for this study, although we believe that VisIt or other tools that have offline Python scripting capability could work equally well. We use the Python API of OpenAI GPT-4 as the LLM.

Because GPT-4 is not trained on the intricacies of data analysis and visualization tool chains, we find that simply invoking the chat interface to GPT-4 does not produce correct code. As such, additional development was needed to improve the resulting code quality. In this paper, we describe the development of ChatVis and demonstrate its use in generating several canonical visualization pipelines. To our knowledge, this is the first such use of LLM synthetic software generation for scientific visualization.
The contributions of this paper are:
%\vspace{0.1cm}
\bi
\im A natural language assistant to generate and execute Python scripts for scientific visualization using few-shot prompting
\im An error detection and correction loop that iteratively extracts error messages from executions of the generated script and generates subsequent prompts to the LLM to correct the specified error
\im Automatically generated visualizations from the synthetically generated scripts
\im Comparison of visualizations against ground truth and attempts to generate the same results with GPT-4 without our agent, for five canonical visualization pipelines of varying complexity
\ei

The rest of this paper is organized as follows. Section~\ref{sec:Related} presents related work in coding assistants and an overview of the scientific visualization tools and tasks used in this study. Section~\ref{sec:Methodology} describes the development of our code assistant agent. Section~\ref{sec:Results} details the visualization scripts that were generated by our agent. Section~\ref{sec:Conclusion} concludes the paper with a summary and outlook to the future of code generation for scientific visualization.

%% file: Related.tex
In this section, we review the relevant literature on using LLMs to generate scripts for scientific visualization. We have organized the literature into two categories: AI coding assistants for applications other than scientific visualization, and a very brief overview of the scientific visualization tools and techniques used in this paper.

\subsection{Coding Assistants}
MyCrunchGPT~\cite{kumar2023mycrunchgpt} is an assistant for scientific machine learning (SciML) tasks. Users interact with MyCrunchGPT over a web-based GUI for various SciML tasks such as training a physics-informed neural network to solve a partial differential equation. The authors use prompting techniques to achieve this. The LM4HPC~\cite{chen2023lm4hpc} framework aims to facilitate the use of language models for HPC specific tasks. It is built on top of several components in the ML software stack with a HuggingFace compatible API. It uses LangChain to improve model performance by integrating new HPC-specific data. It also uses its specific tokenizer, LM4HPC tokenizer, to tackle the input size limit for tokens as HPC tasks can involve scientific codes with large codebases. HPC-GPT~\cite{ding2023hpc} is based on LLaMA and fine-tuned specifically for the HPC domain using generated question-and-answer instances. It has been used for two HPC tasks: for retrieving models and datasets for HPC, and for data race detection in OpenMP programs. Finally, HPC-Coder~\cite{nichols2024hpc} is an LLM fine-tuned to model HPC and scientific codes and used for OpenMP pragma labeling and code generation. In particular, the authors use the HPC source code available on GitHub repositories to fine-tune their model. 

\subsection{Scientific Visualization}

Although hand-drawn visualizations of scientific data predate computer graphics by at least 100 years, modern scientific visualization as a computer science discipline is arguably 37 years old, beginning in 1987 with the seminal report from a workshop of the National Science Foundation~\cite{mccormick88}. Many textbooks provide thorough coverage of the main visualization algorithms~\cite{nielson97, hansen05, bethel12}. Over the years, mature visualization tools---both open-source and commercial---with sophisticated user interfaces have evolved to enable the visualization of scientific datasets. In the open-source domain, ParaView~\cite{ayachit15} and VisIt~\cite{childs12} are commonly used. We used ParaView for this research; VisIt has different syntax for its scripting and user interfaces but otherwise provides similar overall functionality. These tools allow the user to chain together various visualization ``filters,'' where each filter corresponds to a different analysis or visualization operation. The filters are implemented in VTK~\cite{schroeder98}, an open-source software library containing many of the visualization algorithms developed by the scientific visualization community over the last three decades. 

The following filters are featured in our experiments. (1) Contouring is used to extract regions of data with the same scalar value. In 3-d, contouring is called isosurfacing. Contouring and isosurfacing are performed with the marching squares algorithm in 2-d and marching cubes algorithm in 3-d~\cite{lorensen88}. (2) Slicing and clipping are used to extract regions of data on a plane (slicing) or to one side of a plane (clipping). A variation of marching cubes is used for these operations. (3) Volume rendering generates a rendering of a 3-d dataset with color and opacity custom-tailored to the data values, allowing the interior regions of a volume to become visible. Much literature has been published on volume rendering algorithms~\cite{kaufman05}. (4) Delaunay triangulation is used to convert a set of unstructured points into a simplicial mesh, with a number of guaranteed properties of the shape of the resulting simplices~\cite{lawson86}. (5) Streamline tracing is one of several algorithms for visualizing a field of vector-valued data, commonly called flow visualization~\cite{weiskopf05}. In addition to the filters described above, no visualization would be complete without defining view parameters that describe the position of the viewer, the direction being viewed, and the angle of the field of view. The view parameters are used to control the projection of 3-d data into a 2-d image.

%To the best of our knowledge, to date there is no other published work in using LLMs to synthetically generate scientific visualization pipelines. 
In conclusion, while there is a substantial body of work surrounding the tools and techniques of scientific visualization, and considerable advancements in the use of LLMs for tasks within scientific machine learning and high-performance computing, the integration of these two fields remains underexplored. 
Specifically, there appears to be a significant gap in literature concerning the use of LLMs to synthetically generate complete scientific visualization pipelines. To the best of our knowledge, no published work to date directly addresses this particular application.

%% file: Methodology.tex
\begin{figure}
\hspace{-0.19in}
%    \begin{center}
\includegraphics[width=8.9cm]{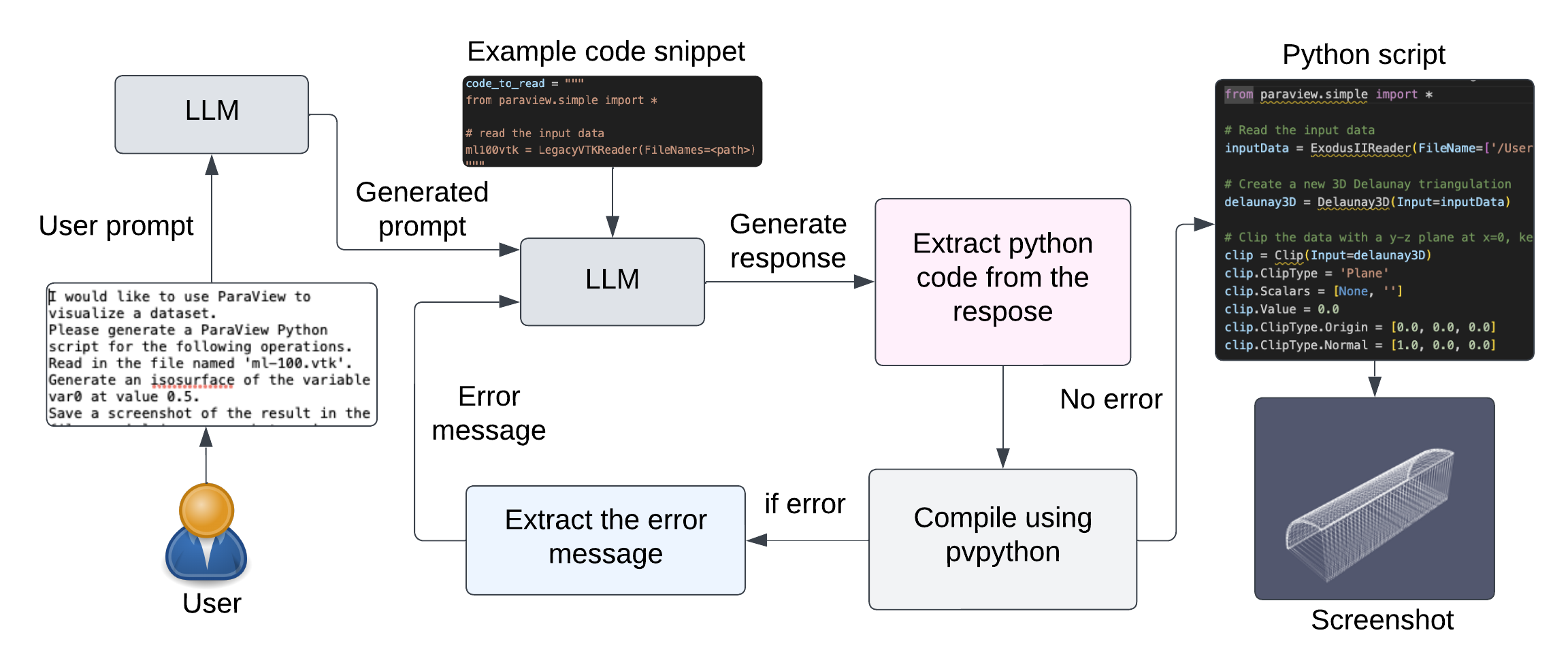}
%    \end{center}
    \vspace{-0.25in}
    \caption{Logical flow of ChatVis execution. User input is processed by an LLM to create a specialized prompt, which, alongside example code snippets, helps the LLM generate a script. This script is executed and any error messages are fed back into the LLM for iterative improvement until a 2D visualization is successfully produced.}
    \label{fig_overview}
    \vspace{-0.25in}
\end{figure}

In this section we describe our methodology to generate an accurate ParaView Python script from user input written in natural language. We propose an LLM-based method as outlined in Figure \ref{fig_overview}. Initially, the user describes their visualization needs in natural language, and an LLM processes this input to generate a more effective prompt. This prompt, combined with multiple example code snippets, is used by the LLM to create a Python script. We then execute this script using ParaView’s PvPython API. If no errors are detected, the script produces a 2D screenshot. If errors occur during script execution, the error messages are fed back to the LLM for corrections, creating a feedback loop that continuously refines the script. This iterative process allows for ongoing improvements based on the error messages until an error-free script is achieved. Upon successful execution, this refinement process generates an error-free Python script and a screenshot of the visualization for evaluation.

\subsection{User input and prompt generation}

Initially, users provide their visualization requirements in natural language, describing what they wish to visualize along with any specific directives or desired features. Upon receiving this input, we employ a language model to process the information and generate a prompt more suitable for scripting with ParaView that maintains the user's specifications. Our methodology involves feeding the LLM both the user's input and a previously crafted example prompt. By analyzing the structure and detailed requirements of the example input and its corresponding prompt, the LLM generates a new prompt that identifies the operations mentioned by the user and arranges them in a step-by-step manner that helps to break down the complex request into smaller, sequential steps to generate high-quality code. This generated prompt includes tasks such as file reading, filter operations, element rendering, setting the camera position, and capturing screenshots at specified resolutions. %This process leverages the LLM’s ability to interpret context and details from minimal examples, effectively refining and translating the natural language input into a structured directive that can be directly utilized for scripting.

\subsection{Script generation using few-shot prompting}
%\tm{COT}
After generating the prompt with step-by-step instructions for creating the Python script, we feed the generated prompt along with a set of example function calls for various operations into the LLM to generate the Python script. We utilize chain-of-thought \cite{wei2022chain} prompting, where the LLM is guided through the logical progression needed to address a complex task. This step-by-step logical breakdown approach helps mitigate common issues where LLMs generate function calls in random order instead of executing tasks logically according to user requirements. Providing examples also helps the LLM avoid generating function calls that do not exist in the ParaView library, thereby preventing syntax errors.
With this structured input and examples, the language model can successfully generate a correct and functional script for the intended visualization tasks.
These examples encompass reading input data and configuring visualization filters like slices, contours, clips, glyphs, tubes, and stream tracers. %Each script specifies properties for these filters, such as setting origins, normals, contour levels, and display colors. 
The scripts also detail how to manage render views by setting view sizes and directions, applying isometric views, creating layouts, displaying visualization data, and saving screenshots. These steps are crucial for generating a correct script.

\subsection{Error detection and correction loop}
After we generate the ParaView Python script, we execute it using ParaView's PvPython API. However, if there are any errors during execution, ParaView will fail to produce a visualization. To handle this, we first developed a tool to detect and extract error messages from the PvPython output, which may include warnings and errors, among other system messages. This tool operates by first splitting the output into individual lines and initializing a list to store these messages. It then identifies tracebacks, which typically start with \texttt{File}, and gathers subsequent lines until it encounters specific errors, such as \texttt{AttributeError}. Once all relevant lines are collected, the function compiles these into a list and returns the error messages. The extracted error messages are sent to the language model with a prompt to fix the code and generate the visualization.
%The extracted error messages are sent to the language model with a prompt stating, "We encountered a Python error: \{error message\}. Can you fix the code \{script\} to generate the visualization based on the given prompt \{prompt\}?". 
The LLM uses this feedback to modify the script.
The revised script then undergoes the same cycle of execution, error message extraction, and script modification using the LLM  until an error-free version is achieved. This iterative process ensures that the generated ParaView Python script is error-free. 

\subsection{Visualization output and evaluation}
We execute the final version of the script using PvPython and generate a screenshot to visualize the output. For evaluation, we also manually perform the same operations using the ParaView GUI, saving both the final Python script and a screenshot. This facilitates a thorough comparison between the LLM-generated script and the manually created script. We assess both the scripts and the screenshots to ensure that the automated script faithfully replicates the manual process in terms of accuracy and visual fidelity. This comparison is crucial as it verifies that the LLM has correctly interpreted the visualization requirements and translated them into a script that produces the desired output.

%% file: Results.tex
We evaluate our approach through a series of common visualization tasks including isosurfacing, slicing, volume rendering, triangulation, streamline tracing, and view rendering.  We conducted our experiments on a workstation equipped with an Apple M2 Max chip with 12‑core CPU, 30‑core GPU, and 96~GB unified memory. Our iterative few shot prompting-based assistant uses the GPT-4 language model from OpenAI~\cite{achiam2023gpt}, where the GPT-4 engine is the latest version of ChatGPT with 1.7 trillion parameters. We compare the images and scripts produced by ChatVis against those produced by manual construction and unassisted prompts to GPT-4. 
We use PvPython version 5.12.0 to generate these images.

\subsection{Isosurfacing}

The first visualization task we consider is to generate an isosurface of a 3-d volume. We use the synthetic Marschner-Lobb benchmark dataset~\cite{marschner94} as the input data. The set of operations consists of reading in the dataset, creating an isosurface where the data have a given value, and saving an image of the isosurface. The prompt provided by the user is shown below.

%User prompt
%backgroundcolor=cyan!20
\begin{myenv}{User prompt}
Please generate a ParaView Python script for the following operations.
Read in the file named ml-100.vtk.
Generate an isosurface of the variable var0 at value 0.5.
Save a screenshot of the result in the filename ml-iso-screenshot.png.
The rendered view and saved screenshot should be 1920 x 1080 pixels.
\end{myenv}

Given an example and the user prompt, we use an LLM to generate a step-by-step instruction prompt for creating the final script. These step-by-step instructions help to outline the operations to be performed sequentially, which aids the LLM in generating the final script accurately. We follow this procedure for all our experiments, but due to space constraints, we report only one example. The LLM-generated prompt corresponding to the user prompt is shown below.

\begin{myenv}{Generated prompt}
Generate a Python script using ParaView for performing visualization tasks based on the provided steps. This script utilizes ParaView to visualize an isosurface from the ml-100.vtk file. Operations include reading the file, generating an isosurface, setting the view resolution, and saving a screenshot. Requirements step-by-step:
\begin{itemize}[leftmargin=*]
    \item Read the file ml-100.vtk given the path.
    \item Generate an isosurface of the variable var0 at value 0.5.
    \item Configure the rendered view resolution to 1920 x 1080 pixels.
    \item Save a screenshot of the rendered view to ml-iso.png.
\end{itemize}
\end{myenv}

\begin{figure}[t]
% \centering
 \hspace{-0.5in}
 \subfigure[Ground truth]{
%   \centering
\includegraphics[width=0.16\textwidth]{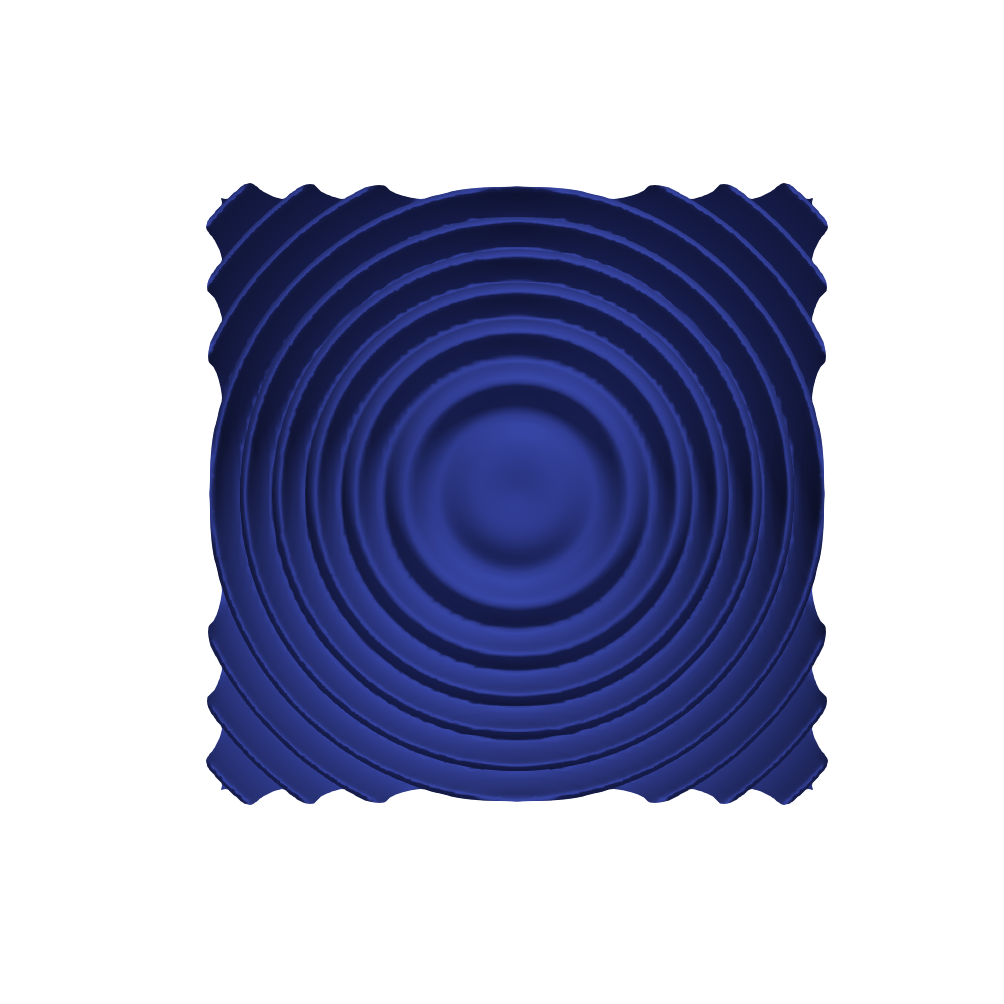} 
   \label{fig:iso-original}
 }
 \hspace{-0.30in}
 \subfigure[ChatVis]{
% \centering
\includegraphics[width=0.16\textwidth]{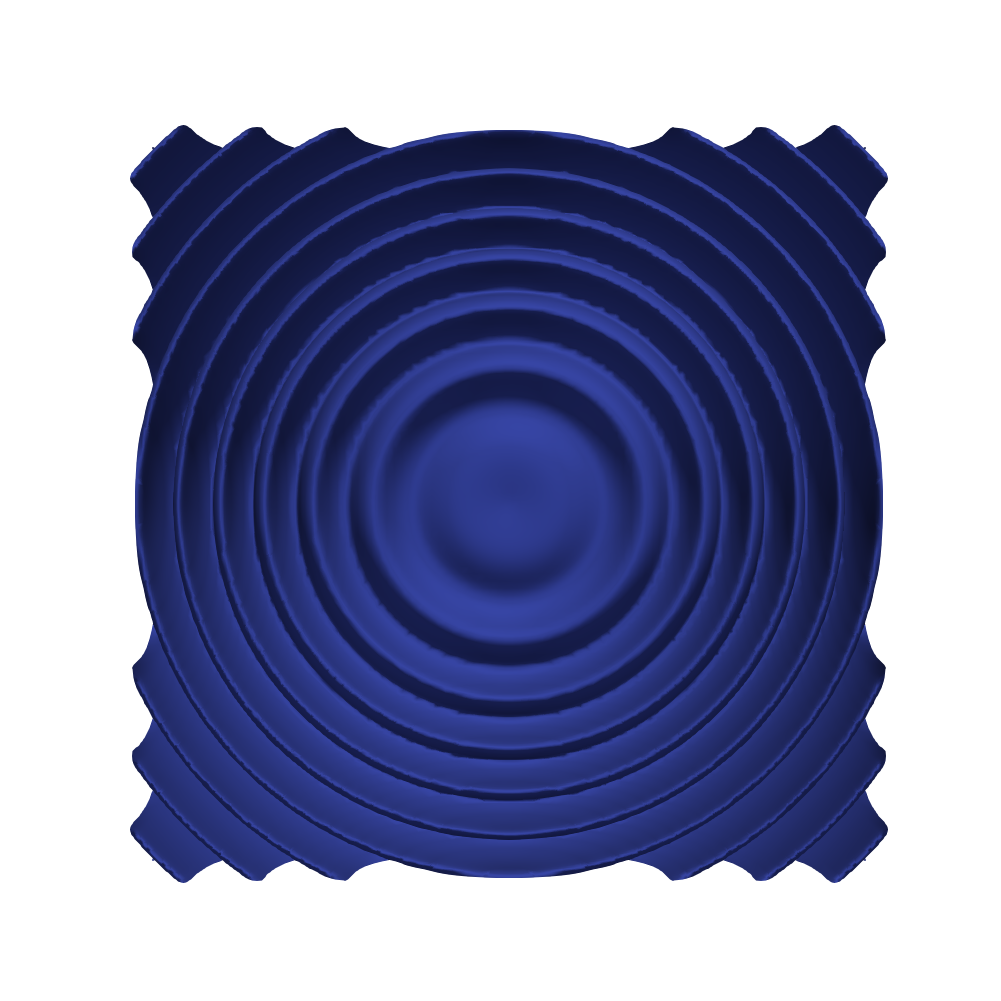}
 \label{fig:iso-chatvis}
 }
 \hspace{-0.1in}
  \subfigure[GPT-4]{
% \centering
\includegraphics[width=0.16\textwidth]{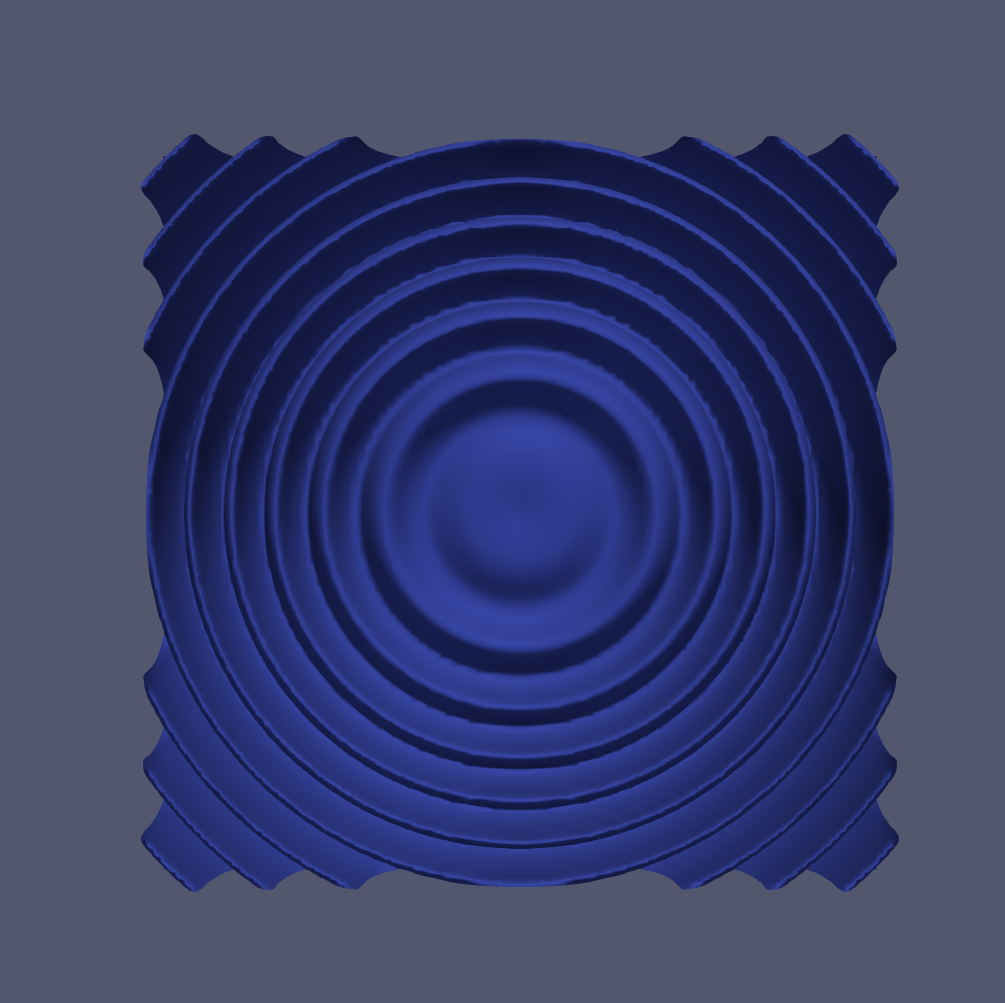}
 \label{fig:iso-gpt4}
 }
 \vspace{-0.1in}
 \caption{Generated images for isosurfacing. } 
 \label{fig:iso}
\vspace{-0.1in}
\end{figure}

%Generated figures
Figure~\ref{fig:iso} displays the generated images for the isosurfacing task. Figure~\ref{fig:iso-original} is generated through manual use of the ParaView GUI and represents our ``ground truth,'' while Figure~\ref{fig:iso-chatvis} is generated by ChatVis, and Figure~\ref{fig:iso-gpt4} is generated by GPT-4. We observe that both ChatVis and GPT-4 can correctly perform the requested operations. This is the only example where GPT-4 produces a correct image as this is a relatively simple task. GPT-4 created a gray background, which is different from the default ParaView script. Meanwhile, the ChatVis background matched the ground truth because ChatVis learned to specify the background color to be white. The camera position was also not specified by the user, leading to slightly different default zoom levels. Later experiments specify camera parameters.

\subsection{Slicing followed by contouring}

The second visualization task takes a slice of the same volumetric dataset as the previous task, and follows with an isocontour at a given value. This demonstrates the ability to link filters together in a pipeline, such that the output of one filter becomes the input to the next. Color mapping is specified, and the view direction is also rotated from the default view, demonstrating simple view manipulation. The prompt provided by the user is shown below.

\begin{myenv}{User prompt}
Please generate a ParaView Python script for the following operations.
Read in the file named `ml-100.vtk'.
Slice the volume in a plane parallel to the y-z plane at x=0.
Take a contour through the slice at the value 0.5.
Color the contour red.
Rotate the view to look at the +x direction.
Save a screenshot of the result in the filename `ml-slice-iso-screenshot.png'.
The rendered view and saved screenshot should be 1920 x 1080 pixels.
\end{myenv}

\begin{figure}[t]
\vspace{-0.1in}
% \centering
\hspace{-0.15in}
 \subfigure[Ground truth]{
%   \centering
% \fbox{
\includegraphics[width=0.22\textwidth]{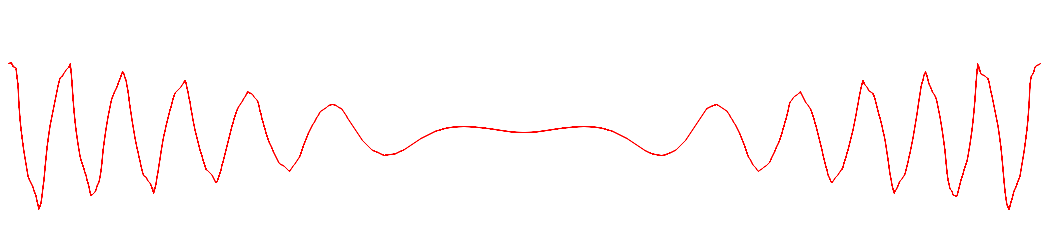} }
   \label{fig:slice-iso-original}
% }
%\hspace{-0.1in}
 \subfigure[ChatVis]{
% \centering
% \fbox{
\includegraphics[width=0.22\textwidth]{figs/ml-slice-iso-screenshot-chatvis-crop.png}}
 \label{fig:slice-iso-chatvis}
% }
 \vspace{-0.1in}
 \caption{Generated images for slicing followed by contouring.} 
 \label{fig:slice-iso}
\vspace{-0.1in}
\end{figure}

Figure \ref{fig:slice-iso} compares the ground truth screenshot with the one generated by ChatVis. ChatVis executed all operations correctly and produced a screenshot identical to the ground truth. 
On the other hand, the code generated by GPT-4 encountered syntax errors because the script attempted to access non-existent attributes. 
Specifically, it attempted to access the \texttt{UseSeparateColorMap} attribute of the \texttt{Contour} class by calling \texttt{ColorBy(contour, None)} and to set the \texttt{ViewUp} attribute on the \texttt{RenderView} class using the code \texttt{view.ViewUp = [0.0, 1.0, 0.0]}. %\tm{All the code should be texttt}

\subsection{Volume rendering}

The third visualization task performs volume rendering on the same dataset as the previous task. This example demonstrates the ability to generate a default color and opacity transfer function based on the extent of the values in the data. The view direction is also rotated to an isometric view, demonstrating more complex view manipulation. The prompt provided by the user is shown below.

\begin{myenv}{User prompt}
Please generate a ParaView Python script for the following operations.
Read in the file named `ml-100.vtk'.
Generate a volume rendering using the default transfer function.
Rotate the view to an isometric direction.
Save a screenshot of the result in the filename `ml-dvr-screenshot.png'.
The rendered view and saved screenshot should be 1920 x 1080 pixels.
\end{myenv}

\begin{figure}[b]
 \vspace{-0.3in}
 \centering
 \subfigure[Ground truth]{
   \centering
   \includegraphics[width=0.18\textwidth]{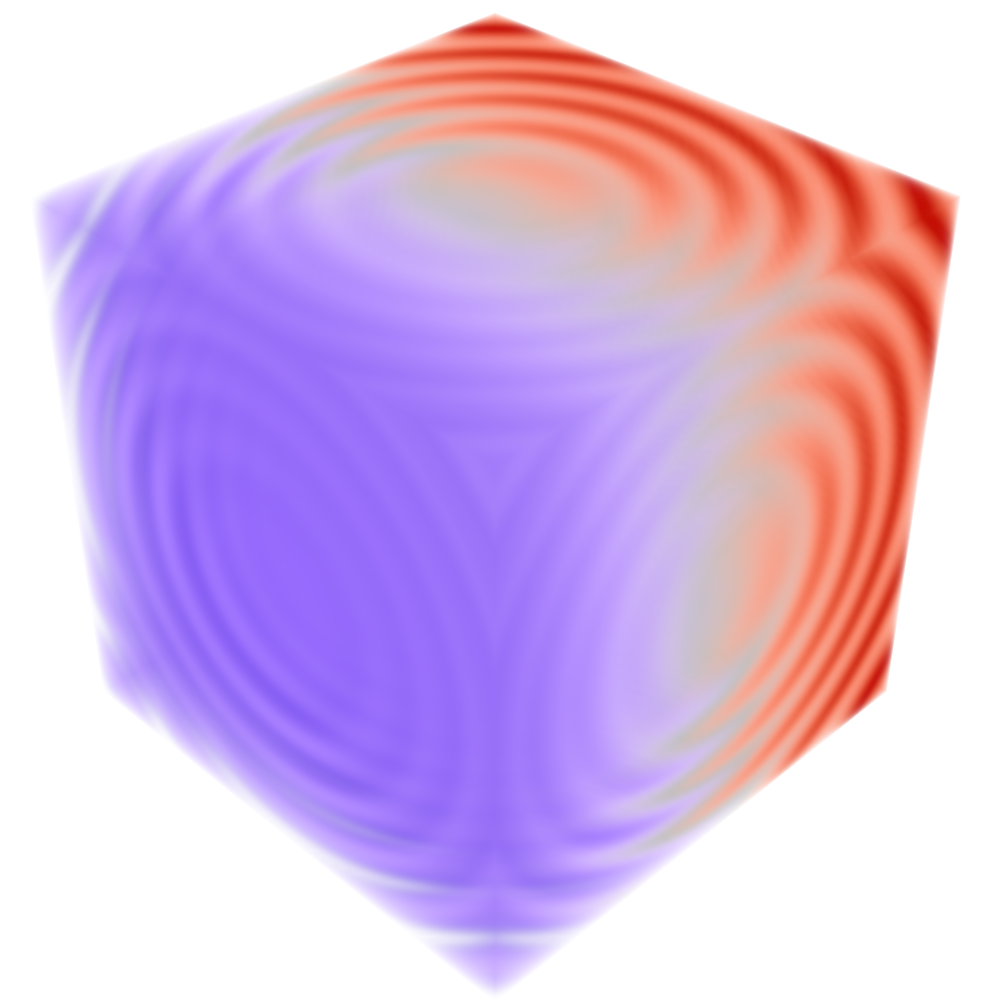} 
   \label{fig:dvr-original}
 }
 \subfigure[ChatVis]{
 \centering
 \includegraphics[width=0.18\textwidth]{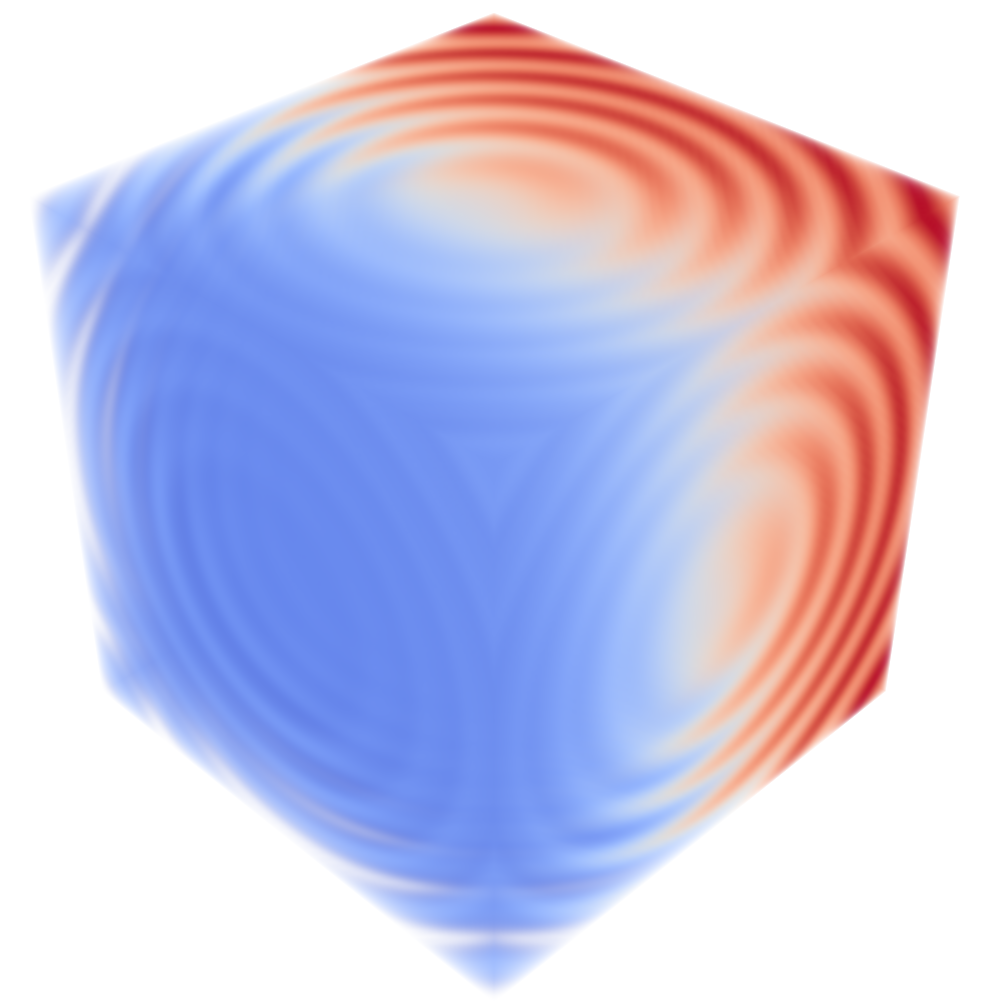}
 \label{fig:dvr-chatvis}
 }
 \vspace{-0.1in}
 \caption{Generated images for volume rendering.} 
 \label{fig:dvr}
            \vspace{-0.15in}
\end{figure}

Figure \ref{fig:dvr} compares the ground truth screenshot with the one generated by ChatVis. ChatVis executed all operations correctly and produced a screenshot identical to the ground truth, except for a different color palette because the user prompt did not specify one. Conversely, the code generated by GPT-4 did not create a volume rendering of the data. Although the code did not produce any errors, it resulted in a blank screenshot because the generated script did not include a volume rendering command. %\tm{Orcun: Tanwi, I can't find the insideOut in the GPT-4 code below.} 
%clipFilter.InsideOut = True

\subsection{Delaunay triangulation}

The fourth visualization task converts a point dataset to a surface through Delaunay triangulation. The input dataset is a point cloud extracted manually from sample data provided with ParaView. This point cloud is read and triangulated. The mesh is then clipped by a plane, keeping only one half of the resulting triangles, further demonstrating the linking of filters together. Rendering is performed in wireframe style as an additional test of various rendering options.

\begin{myenv}{User prompt}
Please generate a ParaView Python script for the following operations.
Read in the file named `can\_points.ex2'.
Generate a 3d Delaunay triangulation of the dataset.
Clip the data with a y-z plane at x=0, keeping the -x half of the data and removing the +x half.
Render the image as a wireframe.
View the result in an isometric view.
Save a screenshot of the result in the filename `points-surf-clip-screenshot.png'.
The rendered view and saved screenshot should be 1920 x 1080 pixels.
\end{myenv}

\begin{figure}[t]
\vspace{-0.1in}
 \centering
 \subfigure[Ground truth]{
   \centering
   \includegraphics[width=0.18\textwidth]{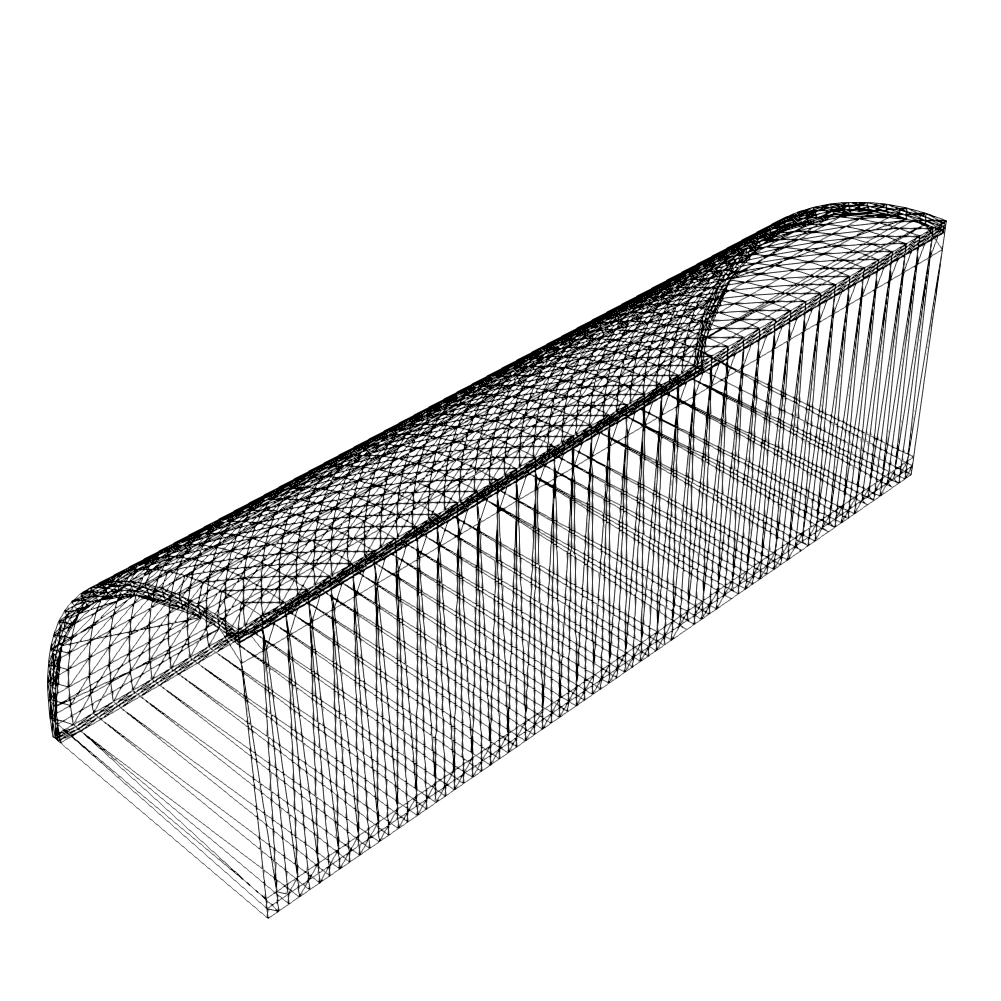} 
   \label{fig:clip-original}
 }
 \subfigure[ChatVis]{
 \centering
 \includegraphics[width=0.18\textwidth]{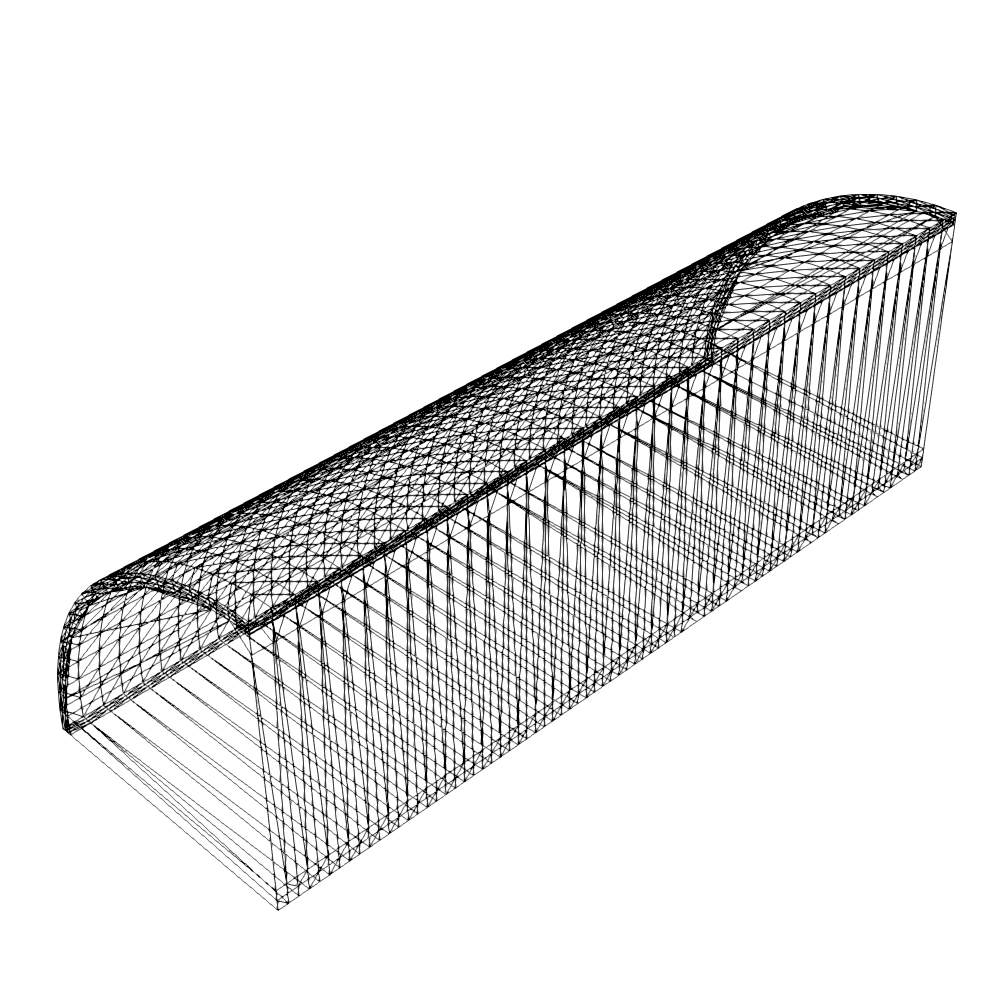}
 \label{fig:clip-chatvis}
 }
 \vspace{-0.1in}
 \caption{Generated images for Delaunay triangulation.} 
 \label{fig:clip}
 \vspace{-0.2in}
\end{figure}

Figure~\ref{fig:clip} shows the generated images for the Delaunay triangulation task, where Figure~\ref{fig:clip-original} is the ground truth, and Figure~\ref{fig:clip-chatvis} is generated by ChatVis. When we compare the ground truth and the image ChatVis generated, we can see that that we were able to correctly perform Delaunay triangulation.
GPT-4 failed to generate an image because the script incorrectly assigned the \texttt{InsideOut} attribute to \texttt{clipFilter}, which does not have an \texttt{InsideOut} method.

 \vspace{-0.1in}
\subsection{Streamline tracing}

The fifth visualization task generates a set of streamlines from a vector-valued flow field. The input dataset is taken from sample data provided with ParaView. The flow velocity field and other scalar fields are read in and streamlines are traced through the velocity data. The streamlines originate at default seed positions. The streamlines are rendered using 3d tubes, and arrow-like glyphs are added to indicate flow direction. Streamlines and glyphs are color-mapped to a scalar data field, in this case temperature.

\begin{myenv}{User prompt}
Please generate a ParaView Python script for the following operations.
Read in the file named `disk.ex2'.
Trace streamlines of the V data array seeded from a default point cloud.
Render the streamlines with tubes.
Add cone glyphs to the streamlines.
Color the streamlines and glyphs by the Temp data array.
View the result in the +X direction.
Save a screenshot of the result in the filename `stream-glyph-screenshot.png'.
The rendered view and saved screenshot should be 1920 x 1080 pixels.
\end{myenv}

\begin{figure}[t]
\vspace{-0.1in}
 \centering
 \subfigure[Ground truth]{
   \centering
   \includegraphics[width=0.18\textwidth]{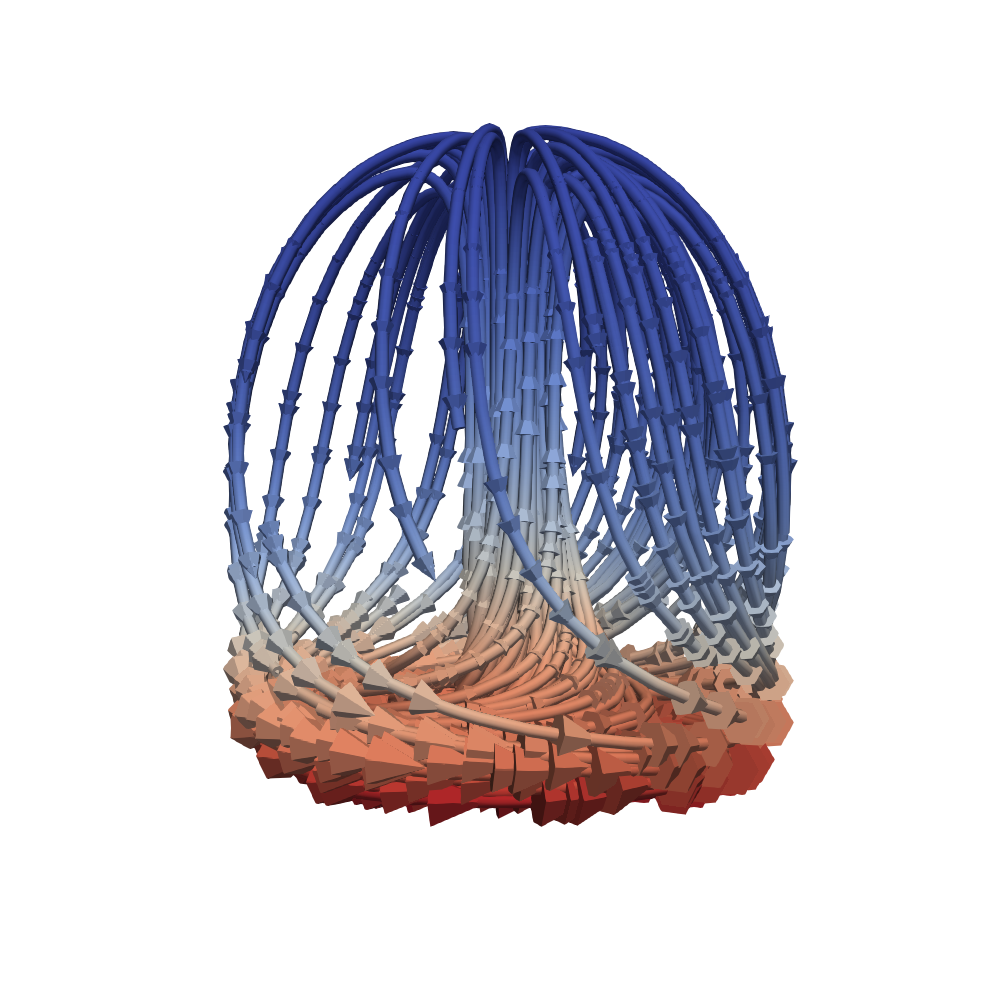} 
   \label{fig:streamline-original}
 }
 \subfigure[ChatVis]{
 \centering
 \includegraphics[width=0.18\textwidth]{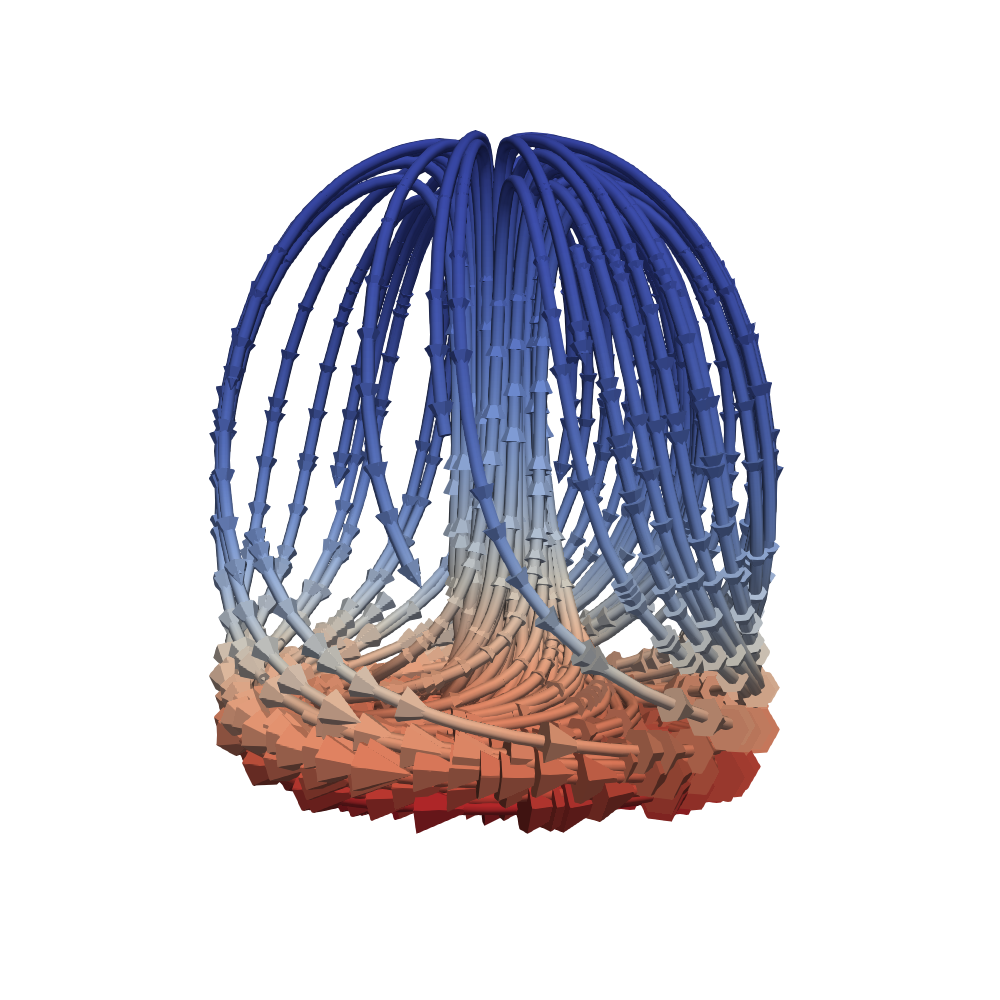}
 \label{fig:streamline-chatvis}
 }
\vspace{-0.1in}
\caption{Generated images for streamline tracing.} 
 \label{fig:streamline}
\vspace{-0.2in}
\end{figure}

%Generated figures
Figure~\ref{fig:streamline} shows the generated images for the streamline tracing task, where Figure~\ref{fig:streamline-original} is the ground truth, and Figure~\ref{fig:streamline-chatvis} is generated by ChatVis. Upon comparing the ground truth image and the image ChatVis generated, we are unable to detect any differences, indicating that ChatVis successfully generated an accurate script. 
%GPT-4 generated an incorrect script that erroneously attributed a \texttt{Scalars} property to the Glyph class, indicating a possible typo, outdated method, or incorrect API usage. \tm{refer the table 1}
 
\setbox0=\hbox{%
\begin{minipage}{.48\textwidth}
\begin{lstlisting}[
basicstyle={\tiny\ttfamily},
identifierstyle={\color{black}},
tabsize=2,
language={python},
numbersep=8pt,
numbers=left,
xleftmargin=0.5cm,frame=tlbr,framesep=2pt,framerule=0pt
]               
from paraview.simple import *

# Reading the disk.ex2 file
reader = ExodusIIReader(FileName='disk.ex2')
reader.UpdatePipeline() 

# Tracing streamlines of the V data array seeded from a default point cloud
streamTracer = StreamTracer(registrationName='StreamTracer1', Input=reader, 
                            SeedType='Point Cloud')
#streamTracer.Vectors = ['POINTS', 'V']

# Rendering the streamlines with tubes for better visibility
tube = Tube(registrationName='Tube1', Input=streamTracer)
tube.Radius = 0.075

# Adding cone glyphs to the streamlines to indicate direction
glyph = Glyph(registrationName='Glyph1', Input=streamTracer, GlyphType='Cone')
glyph.OrientationArray = ['POINTS', 'V']
glyph.ScaleArray = ['POINTS', 'V']
glyph.ScaleFactor = 0.05

# Create a new view and set its properties
renderView = CreateView('RenderView')
renderView.ViewSize = [1920, 1080]

# Create a new layout object
layout = CreateLayout(name='Layout')
layout.AssignView(0, renderView)

# Coloring both the streamlines and glyphs using the Temp data array
tubeDisplay = Show(tube, renderView)
glyphDisplay = Show(glyph, renderView)
ColorBy(tubeDisplay, ('POINTS', 'Temp'))
ColorBy(glyphDisplay, ('POINTS', 'Temp'))
tubeDisplay.RescaleTransferFunctionToDataRange(True)
glyphDisplay.RescaleTransferFunctionToDataRange(True)

# Orienting the view to look from the +X direction
renderView.ResetActiveCameraToPositiveX()
renderView.ResetCamera()

# Save a screenshot of the render view
SaveScreenshot('stream-glyph-screenshot.png', renderView, 
               ImageResolution=[1920, 1080],
               OverrideColorPalette='WhiteBackground')

\end{lstlisting}
\end{minipage}
}
\savestack{\listingA}{\box0}
% #oy: use ^{text}^ for highlighting
\setbox0=\hbox{%
\begin{minipage}{.48\textwidth}
\begin{lstlisting}[
basicstyle={\tiny\ttfamily},
identifierstyle={\color{black}},
tabsize=2,
language={python},
numbersep=8pt,
numbers=left,
moredelim={[is][\textcolor{red}]{\^}{\^}}, 
xleftmargin=0.5cm,frame=tlbr,framesep=2pt,framerule=0pt
] 
from paraview.simple import *

# Disable automatic camera reset on 'Show'
paraview.simple._DisableFirstRenderCameraReset()

# Read in the file
disk_ex2 = ExodusIIReader(FileName=['disk.ex2'])
disk_ex2.PointVariables = ['V', 'Temp']

# Create a stream tracer
streamTracer = StreamTracer(Input=disk_ex2,
                            SeedType='Point Cloud')
streamTracer.Vectors = ['POINTS', 'V']
streamTracer.SeedType.NumberOfPoints = 100  # using default point cloud

# Render the streamlines with tubes
tube = Tube(Input=streamTracer)
tube.Radius = 0.05

# Add cone glyphs to the streamlines
coneGlyph = Glyph(Input=tube,
                  GlyphType='Cone')
^coneGlyph.Scalars = ['POINTS', 'Temp'] # Not a valid function call^
^coneGlyph.Vectors = ['POINTS', 'V'] # Not a valid function call^ 
coneGlyph.ScaleFactor = 1.0

# Coloring both the streamlines and glyphs by the Temp data array
^tubeRepresentation = Show(tube, 'RenderView1') # used before creating a view^
tubeRepresentation.ColorArrayName = ['POINTS', 'Temp']
tubeRepresentation.LookupTable = GetLookupTableForArray('Temp', 1)

glyphRepresentation = Show(coneGlyph, 'RenderView1')
glyphRepresentation.ColorArrayName = ['POINTS', 'Temp']
glyphRepresentation.LookupTable = GetLookupTableForArray('Temp', 1)

# Get the active view and set the view direction
renderView1 = GetActiveViewOrCreate('RenderView')
renderView1.ViewSize = [1920, 1080]
# looking from +X direction 
renderView1.CameraPosition = [1, 0, 0]  ^# CameraPosition is not right^
renderView1.CameraFocalPoint = [0, 0, 0] ^# CameraFocalPoint is not right^
renderView1.CameraViewUp = [0, 0, 1] ^# CameraViewUp is not right^

# Save a screenshot
SaveScreenshot('stream-glyph-screenshot.png', renderView1,
               ImageResolution=[1920, 1080]) ^# Creates gray background^

# Render the final view
Render()

\end{lstlisting}    
\end{minipage}
}
\savestack{\listingB}{\box0}

\begin{table*}[htbp]
\centering
\begin{tabular}{|c|c|}
\hline
{\listingA} &
{\listingB} \\
\hline
\end{tabular}
\caption{Generated Python scripts with ChatVis (left), and GPT-4 (right) for streamline tracing.} \label{tab:script-streamline}
           \vspace{-0.2in}
\end{table*}

For this example, we also report the generated Python scripts created by ChatVis and GPT-4. These scripts are shown in Table~\ref{tab:script-streamline}. Although GPT-4 provides a similar response to our approach, it generates hallucinations in several places, due to the lack of knowledge for this particular visualization task. For instance, an error arises when the script attempts to set the \texttt{Scalars} and \texttt{Vectors} attribute on a Glyph object, which according to the error message, does not exist: 'AttributeError: type object 'Glyph' has no attribute \texttt{Scalars} and \texttt{Vectors}. Moreover, it used \texttt{RenderView1} on Line 28 before this view was created. Additionally, the camera view and position result in cropped screenshots from different viewing angles. The code generated by ChatVis in lines 39 and 40 effectively resets the camera view and orientation, ensuring that the entire object is captured in the screenshot. Overall, our proposed methodology of breaking down the problem step-by-step and providing ParaView example code snippets helped to ensure the correct function calls were made in the proper order.
%\tm{how ChatVis did it correct from the methodology perspective?}

\subsection{Comparisons with other LLM models}

\begin{table*}[h] %ht!
\centering
\begin{small}

\begin{tabular}{l|ll|ll|ll|ll|ll|ll|}
\cline{2-13}
                                                                                                & \multicolumn{2}{c|}{\textbf{ChatVis}}   & \multicolumn{2}{c|}{\textbf{GPT-4}}     & \multicolumn{2}{c|}{\textbf{GPT-3.5}}   & \multicolumn{2}{c|}{\textbf{Llama 3:8b}} & \multicolumn{2}{c|}{\textbf{Codellama:7}} & \multicolumn{2}{c|}{\textbf{Codegemma}} \\ \hline
\multicolumn{1}{|c|}{\textbf{Visualizations}}                                                   & \multicolumn{1}{l|}{Error} & SS & \multicolumn{1}{l|}{Error} & SS & \multicolumn{1}{l|}{Error} & SS & \multicolumn{1}{l|}{Error}  & SS & \multicolumn{1}{l|}{Error}  & SS  & \multicolumn{1}{l|}{Error} & SS \\ \hline
\multicolumn{1}{|l|}{Isosurfacing}                                                              & \multicolumn{1}{l|}{\cellcolor{green}No}    & \cellcolor{green}Yes        & \multicolumn{1}{l|}{ \cellcolor{green} No}    & \cellcolor{green} Yes        & \multicolumn{1}{l|}{\cellcolor{RED} Yes}   & \cellcolor{RED} No         & \multicolumn{1}{l|}{\cellcolor{RED}Yes}    & \cellcolor{RED}No         & \multicolumn{1}{l|}{\cellcolor{RED}Yes}    & \cellcolor{RED}No          & \multicolumn{1}{l|}{\cellcolor{RED}Yes}   & \cellcolor{RED}No         \\ \hline
\multicolumn{1}{|l|}{\begin{tabular}[c]{@{}l@{}}Slicing then contouring\end{tabular}} & \multicolumn{1}{l|}{\cellcolor{green}No}    & \cellcolor{green}Yes        & \multicolumn{1}{l|}{\cellcolor{RED}Yes}   & \cellcolor{RED}No         & \multicolumn{1}{l|}{\cellcolor{RED}Yes}   & \cellcolor{RED}No         & \multicolumn{1}{l|}{\cellcolor{RED}Yes}    & \cellcolor{RED}No         & \multicolumn{1}{l|}{\cellcolor{RED}Yes}    & \cellcolor{RED}No          & \multicolumn{1}{l|}{\cellcolor{RED}Yes}   & \cellcolor{RED}No         \\ \hline
\multicolumn{1}{|l|}{Volume rendering}                                                          & \multicolumn{1}{l|}{\cellcolor{green}No}    & \cellcolor{green}Yes        & \multicolumn{1}{l|}{\cellcolor{green}No}    & \cellcolor{RED}No         & \multicolumn{1}{l|}{\cellcolor{RED}Yes}   & \cellcolor{RED}No         & \multicolumn{1}{l|}{\cellcolor{RED}Yes}    & \cellcolor{RED}No         & \multicolumn{1}{l|}{\cellcolor{RED}Yes}    & \cellcolor{RED}No          & \multicolumn{1}{l|}{\cellcolor{RED}Yes}   & \cellcolor{RED}No         \\ \hline
\multicolumn{1}{|l|}{Delaunay triangulation}                                                    & \multicolumn{1}{l|}{\cellcolor{green}No}    & \cellcolor{green}Yes        & \multicolumn{1}{l|}{\cellcolor{RED}Yes}   & \cellcolor{RED}No         & \multicolumn{1}{l|}{\cellcolor{RED}Yes}   & \cellcolor{RED}No         & \multicolumn{1}{l|}{\cellcolor{RED}Yes}    & \cellcolor{RED}No         & \multicolumn{1}{l|}{\cellcolor{RED}Yes}    & \cellcolor{RED}No          & \multicolumn{1}{l|}{\cellcolor{RED}Yes}   & \cellcolor{RED}No         \\ \hline
\multicolumn{1}{|l|}{Streamline tracing}                                                        & \multicolumn{1}{l|}{\cellcolor{green}No}    & \cellcolor{green}Yes        & \multicolumn{1}{l|}{\cellcolor{RED}Yes}   & \cellcolor{RED}No         & \multicolumn{1}{l|}{\cellcolor{RED}Yes}   & \cellcolor{RED}No         & \multicolumn{1}{l|}{\cellcolor{RED}Yes}    & \cellcolor{RED}No         & \multicolumn{1}{l|}{\cellcolor{RED}Yes}    & \cellcolor{RED}No          & \multicolumn{1}{l|}{\cellcolor{RED}Yes}   & \cellcolor{RED}No         \\ \hline
\end{tabular}
\end{small}
\caption{Performance comparison of various LLMs based on two criterias: (i) whether the model can generate scripts without syntax errors, and (ii) whether the scripts can successfully produce screenshots (SS).}
\label{tab_comp}
           \vspace{-0.1in}
\end{table*}

In addition to GPT-4, we also compared the performance of ChatVis with other state-of-the-art LLMs such as GPT-3.5-turbo, LLaMA-3.8B~\cite{touvron2023llama}, and other models specifically designed for code generation, such as Code LLaMA~\cite{roziere2023code} and Codegemma~\cite{team2024codegemma}. The comparison is based on two criteria: (1) whether the model can generate scripts without syntax errors, and (2) whether the scripts, once executed, can successfully produce a screenshot. Table~\ref{tab_comp} summarizes our findings. In sum, every model except GPT-4 failed to generate scripts without syntax errors, and therefore could not produce screenshots for any of the visualization tasks. As we report above, GPT-4 was only able to generate a correct isosurfacing screenshot, and a script without syntax errors for volume rendering; however, since the script did not perform volume rendering correctly, it could not generate the correct screenshot. This demonstrates that while LLMs excel at creating basic Python scripts for general tasks~\cite{jiang2024survey}, they often fail to generate accurate scripts for specialized tasks such as scientific visualization. However, among all the LLMs evaluated, GPT-4 stood out by generating the least hallucinated code and accurately executing the isosurfacing task. Ultimately, this was the reason we selected GPT-4 as the base comparison model for ChatVis.

           \vspace{-0.1in}

%% file: Conclusion.tex
\vspace{-0.07in}
In this paper, we describe a methodology for generating accurate scientific visualization scripts from natural language inputs. We developed ChatVis, which first processes these inputs to create effective prompts. These prompts, supplemented with example code snippets, guide the LLM in scripting tasks that are compiled using ParaView’s PvPython API. The process incorporates a feedback loop for error correction, enhancing script accuracy through iterative refinements until an error-free script is achieved. This method not only successfully generates both Python scripts and visualization screenshots but also demonstrates the effectiveness of integrating LLMs with domain-specific scripting tasks. On the other hand, the scripts generated by GPT-4 and other LLM models using basic user prompts often hallucinate, creating scripts with non-existent function calls, leading to compilation failures.
% We compared our method against human-generated ground truth, demonstrating that ChatVis can accurately generate correct code and visualizations. Additionally, we evaluated the scripts generated by GPT-4 and other models using basic user prompts and found that it often hallucinates, creating code with non-existent function calls, leading to compilation failures.
In conclusion, ChatVis shows significant potential in accurately automating the generation of complex visualization scripts. This research sets the groundwork for future enhancements and broader applications of machine learning in scientific visualization.

% \subsection{Future Work}
% LLMs are good at producing basic Python scripts for general purposes. However, when tasked with creating specific scripts for specialized software like ParaView, they sometimes generate incorrect or non-existent function calls. To address this issue, 

In the future, we plan to refine the capabilities of ChatVis by fine-tuning it with function calls from ParaView's source code. This targeted approach will improve the accuracy and reliability of script generation for specialized visualization tasks.
Moreover, we will implement automated script evaluation, focusing on assessing the accuracy of the generated code, even without visual output. By systematically analyzing how closely the code matches expected outputs, we can effectively gauge script performance and conduct large-scale evaluations. 
%in addition to evaluating the script through screenshots
%This approach will enable us to conduct evaluations on a larger scale.

%% file: ai4s-IEEE.bbl
\begin{thebibliography}{10}

\bibitem{achiam2023gpt}
Josh Achiam, Steven Adler, Sandhini Agarwal, Lama Ahmad, Ilge Akkaya, Florencia~Leoni Aleman, Diogo Almeida, Janko Altenschmidt, Sam Altman, Shyamal Anadkat, et~al.
\newblock Gpt-4 technical report.
\newblock {\em arXiv preprint arXiv:2303.08774}, 2023.

\bibitem{ayachit15}
Utkarsh Ayachit.
\newblock {\em {The ParaView Guide: A Parallel Visualization Application}}.
\newblock Kitware, Inc., 2015.

\bibitem{bethel12}
E~Wes Bethel, Hank Childs, and Charles Hansen.
\newblock {\em {High Performance Visualization: Enabling Extreme-Scale Scientific Insight}}.
\newblock CRC Press, 2012.

\bibitem{chen2023lm4hpc}
Le~Chen, Pei-Hung Lin, Tristan Vanderbruggen, Chunhua Liao, Murali Emani, and Bronis de~Supinski.
\newblock Lm4hpc: Towards effective language model application in high-performance computing.
\newblock In {\em International Workshop on OpenMP}, pages 18--33. Springer, 2023.

\bibitem{childs12}
Hank Childs, Eric Brugger, Brad Whitlock, Jeremy Meredith, Sean Ahern, David Pugmire, Kathleen Biagas, Mark Miller, Cyrus Harrison, Gunther~H Weber, et~al.
\newblock {VisIt: An End-User Tool for Visualizing and Analyzing Very Large Data}.
\newblock 2012.

\bibitem{ding2023hpc}
Xianzhong Ding, Le~Chen, Murali Emani, Chunhua Liao, Pei-Hung Lin, Tristan Vanderbruggen, Zhen Xie, Alberto Cerpa, and Wan Du.
\newblock Hpc-gpt: Integrating large language model for high-performance computing.
\newblock In {\em Proceedings of the SC'23 Workshops of The International Conference on High Performance Computing, Network, Storage, and Analysis}, pages 951--960, 2023.

\bibitem{hansen05}
Charles~D Hansen and Chris~R Johnson.
\newblock {\em {The Visualization Handbook}}.
\newblock Academic Press, 2005.

\bibitem{jiang2024survey}
Juyong Jiang, Fan Wang, Jiasi Shen, Sungju Kim, and Sunghun Kim.
\newblock A survey on large language models for code generation.
\newblock {\em arXiv preprint arXiv:2406.00515}, 2024.

\bibitem{kaufman05}
Arie~E Kaufman and Klaus Mueller.
\newblock {Overview of Volume Rendering}.
\newblock {\em {The Visualization Handbook}}, 7:127--174, 2005.

\bibitem{kumar2023mycrunchgpt}
Varun Kumar, Leonard Gleyzer, Adar Kahana, Khemraj Shukla, and George~Em Karniadakis.
\newblock Mycrunchgpt: A llm assisted framework for scientific machine learning.
\newblock {\em Journal of Machine Learning for Modeling and Computing}, 4(4), 2023.

\bibitem{lawson86}
Charles~L Lawson.
\newblock {Properties of N-Dimensional Triangulations}.
\newblock {\em Computer Aided Geometric Design}, 3(4):231--246, 1986.

\bibitem{lorensen88}
William~E Lorensen and Harvey~E Cline.
\newblock {Marching Cubes: A High Resolution 3D Surface Construction Algorithm}.
\newblock In {\em Seminal graphics: pioneering efforts that shaped the field}, pages 347--353. 1998.

\bibitem{marschner94}
Stephen~R Marschner and Richard~J Lobb.
\newblock {An Evaluation of Reconstruction Filters for Volume Rendering}.
\newblock In {\em Proceedings Visualization'94}, pages 100--107. IEEE, 1994.

\bibitem{mccormick88}
Bruce~H McCormick.
\newblock Visualization in scientific computing.
\newblock {\em Acm Sigbio Newsletter}, 10(1):15--21, 1988.

\bibitem{nichols2024hpc}
Daniel Nichols, Aniruddha Marathe, Harshitha Menon, Todd Gamblin, and Abhinav Bhatele.
\newblock Hpc-coder: Modeling parallel programs using large language models.
\newblock In {\em ISC High Performance 2024 Research Paper Proceedings (39th International Conference)}, pages 1--12. Prometeus GmbH, 2024.

\bibitem{nielson97}
Gregory Nielson, Hans Hagen, and Heinrich Muller.
\newblock {Scientific Visualization}.
\newblock Institute of Electrical \& Electronics Engineers, 1997.

\bibitem{roziere2023code}
Baptiste Roziere, Jonas Gehring, Fabian Gloeckle, Sten Sootla, Itai Gat, Xiaoqing~Ellen Tan, Yossi Adi, Jingyu Liu, Tal Remez, J{\'e}r{\'e}my Rapin, et~al.
\newblock Code llama: Open foundation models for code.
\newblock {\em arXiv preprint arXiv:2308.12950}, 2023.

\bibitem{schroeder98}
Will Schroeder, Kenneth~M Martin, and William~E Lorensen.
\newblock {\em {The Visualization Toolkit: An Object-Oriented Approach to 3D Graphics}}.
\newblock Prentice-Hall, Inc., 1998.

\bibitem{team2024codegemma}
CodeGemma Team.
\newblock Codegemma: Open code models based on gemma.
\newblock {\em arXiv preprint arXiv:2406.11409}, 2024.

\bibitem{touvron2023llama}
Hugo Touvron, Thibaut Lavril, Gautier Izacard, Xavier Martinet, Marie-Anne Lachaux, Timoth{\'e}e Lacroix, Baptiste Rozi{\`e}re, Naman Goyal, Eric Hambro, Faisal Azhar, et~al.
\newblock Llama: Open and efficient foundation language models.
\newblock {\em arXiv preprint arXiv:2302.13971}, 2023.

\bibitem{wei2022chain}
Jason Wei, Xuezhi Wang, Dale Schuurmans, Maarten Bosma, Fei Xia, Ed~Chi, Quoc~V Le, Denny Zhou, et~al.
\newblock Chain-of-thought prompting elicits reasoning in large language models.
\newblock {\em Advances in neural information processing systems}, 35:24824--24837, 2022.

\bibitem{weiskopf05}
Daniel Weiskopf and Gordon Erlebacher.
\newblock {Flow Visualization Overview}.
\newblock {\em Handbook of Visualization}, pages 261--278, 2005.

\end{thebibliography}
